\documentclass{eas}
\usepackage{graphicx}
\begin{document}

\title{Constraining GRB as Source for UHE Cosmic Rays through Neutrino Observations} 
\author{Pisin Chen}
\address{Department of Physics and Leung Center for Cosmology and Particle Astrophysics (LeCosPA), National Taiwan University, Taipei, Taiwan 10617 \\ \email{pisinchen@phys.ntu.edu.tw}\\
Kavli Institute for Particle Astrophysics and Cosmology, SLAC National Accelerator Laboratory, Menlo Park, CA 94025, U.S.A. \\}

\runningtitle{GRB as Source for UHE Cosmic Rays}
\begin{abstract}
The origin of ultra-high energy cosmic rays (UHECR) has been widely regarded as one of the major questions in the frontiers of particle astrophysics. Gamma ray bursts (GRB), the most violent explosions in the universe second only to the Big Bang, have been a popular candidate site for UHECR productions. The recent IceCube report on the non-observation of GRB induced neutrinos therefore attracts wide attention. This dilemma requires a resolution: either the assumption of GRB as UHECR accelerator is to be abandoned or the expected GRB induced neutrino yield was wrong. It has been pointed out that IceCube has overestimated the neutrino flux at GRB site by a factor of $\sim 5$. In this paper we point out that, in addition to the issue of neutrino production at source, the neutrino oscillation and the possible neutrino decay during their flight from GRB to Earth should further reduce the detectability of IceCube, which is most sensitive to the muon-neutrino flavor as far as point-source identification is concerned. Specifically, neutrino oscillation will reduce the muon-neutrino flavor ratio from 2/3 per neutrino at GRB source to 1/3 on Earth, while neutrino decay, if exists and under the assumption of normal hierarchy of mass eigenstates, would result in a further reduction of muon-neutrino ratio to 1/8. With these in mind, we note that there have been efforts in recent years in pursuing other type of neutrino telescopes based on Askaryan effect, which can in principle observe and distinguish all three flavors with comparable sensitivities. Such new approach may therefore be complementary to IceCube in shedding more lights on this cosmic accelerator question.
\end{abstract}
\maketitle
\section{Introduction - Neutrino as Cosmic Messenger}

Ultra-high energy cosmic rays (UHECR) with energies beyond $10^{19}$ eV have been observed (HiRes \cite{HiRes2009-UHECR-Stereo}; Auger \cite{Auger2010-EeVSpectrum}; ANITA \cite{ANITA2010-UHECR}). Their origin, however, has not been clear. This challenge has been considered as one of the eleven science questions for the new century by the well publicized white-paper prepared for the US National Research Council by the Turner Committee in 2003 (Turner \cite{Turner2003}). Gamma ray bursts (GRB), the most violent explosions in the universe second only to the Big Bang, has long been considered as a promising candidate site for the ``cosmic accelerator" where UHECRs, primarily protons, are generated. Since proton trajectory can be bent by inter- and intra- galactic magnetic fields on its way to Earth, its incoming angle cannot reveal its origin. It happens that a necessary by-product of UHECR on or near the cosmic accelerator, such as GRB, site are the neutrinos at comparable energy (UHECN). Being charge-neutral and weakly interacting, neutrinos so produced can propagate straight-forwardly to Earth. The detection of such neutrinos would therefore provide a useful means to address the cosmic accelerator puzzle.

There are multiple channels where such conversion can happen. We assume that the UHECRs are primarily protons. First, such protons can interact with the intense GRB background within the GRB fireball, possibly comoving, and neutrinos are generated through photo-pion production. Such interaction is particularly pronounced in the $\Delta$-resonance process:
\begin{eqnarray}
p+\gamma \to \Delta \to n+\pi^+ \to n+ \{\mu^+ +\nu_{\mu}\} 
\to n+\{[e^+ +\nu_{e}+\bar{\nu}_{\mu}]+\nu_{\mu}\}.
\end{eqnarray}
The resulting UHECN spectrum should peak at several hundred TeV (Waxman \& Bahcall \cite{Waxman1997}). Such UHECNs are produced strictly on the GRB site. Second, the UHECR protons may instead collide with the cosmic microwave background (CMB) on their flight to Earth, and turn into UHECNs under the same process in Eq.(1). This is the so-called GZK process (Greisen \cite{Greisen1966}; Zatsepin \& Kuzmin \cite{Zatsepin1966}; Berezinsky \& Zatsepin \cite{Berezinsky1969}). For this channel to happen, it would require a much higher UHECR proton energy since CMB photons are much softer than that of GRB (but may collide head-on). The associated neutrino spectrum has a flat top ranging from $10^{15}$eV to $10^{20}$eV (see Fig. 1). Though such UHECNs are produced offsite from the cosmic accelerator, it generally occur very near the source. For example, at redshift $z\sim 1$ the mean-free-path for a UHECR proton above the GZK threshold energy is $< 6 {\rm Mpc}$. A neutrino so produced would arrive at Earth with an incoming angle that is within 5 arc second around the GRB where the UHECR proton was accelerated. The point-back ability of UHECN to the cosmic accelerator is thus not so compromised even through this channel.


\section{Non-observation of GRB Neutrinos at IceCube}

Giving the importance of the issue as described above, the recent IceCube announcement of the non-observation of GRB neutrinos (Abbasi et al. [IceCube Coll.] \cite{IceCube}; Ahlers, Gonzalez-Garcia \& Halzen \cite{Ahlers2011}) came as a disappointment to many. 
IceCube calculated the expected prompt neutrino spectrum in the internal shock scenario of the fireball model following Guetta et al. (\cite{Guetta2004}), which is based on Waxman \& Bacall (\cite{Waxman1997}). As far as point-source observations is concern, IceCube is most sensitive to muons for two reasons (Karle et al. [IceCube Coll.] \cite{Karle2003}): 1.) Muons allow a very good angular resolution of $0.7^{\circ}$ over a wide range of energy; 2.) The effective volume for muons exceeds the geometric volume of the detector by factors of 10-50, depending on energy. Following IceCube and for the rest of this paper, we shall neglect the contributions from the other two flavors in IceCube's sensitivity to address the GRB connection with cosmic accelerator. 

IceCube's starting point is to express the GRB neutrino flux in terms of its UHECR proton flux:
\begin{equation}
\frac{\mathcal{F}_{\nu}^{\rm IC}}{\mathcal{F}_p}=\frac{1}{8}f_{\pi,b},
\end{equation}
where $f_{\pi,b}\equiv f_{\pi}(E=E_b)$ is the fraction of proton energy carried by the pion as a result of proton-GRB photon collision, with a spectral-break energy $\epsilon_b$ at $\Delta$-resonance:
\begin{equation}
E_b=1.3\times10^{16}\Gamma^2_{2.5}\epsilon^{-1}_{b,{\rm MeV}}{\rm eV},
\end{equation}
where $\Gamma=10^{2.5}\Gamma_{2.5}$ is the Lorentz factor of the bulk flow and $\epsilon_b=1\epsilon_{b, {\rm MeV}}$. On the other hand, the proton flux can be normalized by the gamma-ray flux through 
\begin{equation}
\mathcal{F}_p=\frac{1}{f_e}\mathcal{F}_{\gamma},
\end{equation}
where $f_e$ is the ratio of accelerated proton to electron energies. By 2011 with half of its detector completed, IceCube has reached the sensitivity that is comparable to the expected neutrino flux from GRBs. Based on 117 GRB events, IceCube searched for neutrino emission from these sources over a wide range of energies and emission times, but produced no evidence for such, excluding prevailing models at 90\% CL.

\section{Over-estimation of Neutrino Flux at Source}

It has been pointed out, however, that the IceCube Collaboration might have over-estimated the neutrino flux at GRB source by as much as a factor 5 (Zhuo Li \cite{LiZhuo}). If so, then the IceCube results have not yet ruled out the GRB candidacy for the cosmic accelerator. Here we review Li's argument and correction to the GRB neutrino flux. His main points are as follows.

1. For a flat proton distribution with the index given above and the Band-function parameters, $\alpha=-1$ and $\beta=-2.2$, so chosen by IceCube, the assumption of $f_{\pi}(E)=f_{\pi,b}$ is valid only for protons with energy $E>E_b$. For $E<E_b$, $f_{\pi}$ is no longer a constant but instead proportional to the energy: $f_{\pi}\propto E$. The physical reason for this is simple: the higher energy target photons are harder to find to match with the lower energy protons. Thus by applying constant $f_{\pi}$  to all proton energies, Eq.(1) has overestimated the initial neutrino flux. 

2. Eq.(3.1) has also ignored the suppression of neutrino production at high energies due to the radiative cooling of secondary pions/muons. The synchrotron cooling timescale is shorter than that for the secondary decay, and therefore the neutrino production will be suppressed, when the pion/muon energy is above the cooling energy $E_c$.

Thus the neutrino production is mainly contributed by the primary protons with an energy window $E_b<E<E_c$, which is only a fraction of the total number of GRB accelerated UHE protons. Let the maximum and minimum accelerated proton energies be $E_{max}$ and $E_{min}$, respectively, with ratio $E_{max}/E_{min}\sim 10^9$, and let $E_c/E_b\sim 10^2$. One then finds, for a proton energy spectrum $E^2dn_p/dE\propto E^{2-p}$, with $p\approx 2$, 
\begin{equation}
\mathcal{F}_{\nu}= \frac{\mathcal{F}_{\nu}^{\rm IC}}{\mathcal{F}_p}\int^{E_c}_{E_b}E\frac{dn_p}{dE}dE\approx \mathcal{F}_{\nu}^{\rm IC}\frac{\ln(E_c/E_b)}{\ln(E_{max}/E_{min})}\sim 0.22 \mathcal{F}_{\nu}^{\rm IC}.
\end{equation}
Thus the correction to the IceCube assumption of the neutrino flux is the reduction of $\mathcal{F}_{\nu}^{\rm IC}$ by roughly a factor 5. With this correction, Li found that the IceCube data is still consistent with the assumption of GRB as the origin for UHECR.

It should be reminded that there exist additional corrections to the neutrino flux due to various uncertainties in the fundamental physics parameters such as $f_{\pi}$ and $f_e$, as well as in the GRB and UHECR acceleration models.   

\section{Evolution of Neutrino Flavors in Flight}

Having discussed the conversion factor from UHE protons to neutrinos at GRB site, which is GRB model and cosmic accelerator model dependent, in this section we turn our attention to the impact on the GRB-neutrino connection due to neutrino properties in-flight from GRB to Earth. While the notion of neutrino oscillation is by now well-known and has been incorporated into data analysis, it appears that the implication of possible neutrino decay has thus far not been taken into active consideration. 

\subsection{Neutrino Oscillation}
As we know, neutrinos oscillate among their three different flavors (Fukuda et al. [Super-Kamiokande Collaboration] \cite{SuperK1998}). From Eq.(2.1) we see that every proton-photon interaction will produce one electron neutrino and two muon neutrinos. So the ratio between different neutrino flavors at source, per out-coming neutrino, is 
\begin{equation}
{\rm At\ Source:}\quad\quad\quad f^S_e:f^S_{\mu}:f^S_{\tau}= 1/3:2/3:0.
\end{equation} 
Being produced at cosmic distance, which is much much larger than the neutrino oscillation length, the GRB induced ultra-high energy neutrinos flavors will reach an equilibrium when arriving at Earth with a uniform distribution (see, for example, Wang, Chen, Huang et al. \cite{SHWang2013}): 
\begin{equation}
{\rm \quad In\ Flight:}\quad\quad\quad f^E_e:f^E_{\mu}:f^E_{\tau}= 1/3:1/3:1/3.
\end{equation} 
As mentioned earlier, although IceCube is able to detect all neutrino flavors, in its search for point courses such as neutrino emission from GRB, muon-neutrino has the highest sensitivity and best angular resolution. This, however, necessarily reduces the effective neutrino flux to 1/3 of the total flux emitted from GRB. 

\subsection{Neutrino Decay}
Another major discovery about neutrinos in the last two decades is that neutrinos have mass. This, together with additional assumption of the decay process based on notions beyond the standard model of particle physics, leads to the prediction of neutrino decay, from the higher mass eigenstates to the mass ground state. One can envision at least two possible mass hierarchies: the normal hierarchy, i.e., $m_3\gg m_2> m_1$, and the inverted hierarchy, i.e., $m_2> m_1 \gg m_3$. It has been shown that under normal hierarchy the eventual flavor ratio on Earth, again normalized to number of neutrinos produced at GRB site, is (Beacom, Bell, Hooper, et al. \cite{Beacom2003}, Maltoni \& Winter \cite{Maltoni2008})
\begin{equation}
{\rm \quad Normal\ Hierarchy:}  \quad\quad\quad f^E_{e}:f^E_{\mu}:f^E_{\tau}=2/3:1/8:5/24,
\end{equation}
or 
\begin{equation}
{\rm Inverted\ Hierarchy:}  \quad\quad\quad f^E_{e}:f^E_{\mu}:f^E_{\tau}=0:2/5:3/5.
\end{equation}
So for the IceCube search for GRB emission of neutrinos, which is most sensitive to $\nu_{\mu}$, its detector sensitivity would be further reduced from 1/3 (based on pure oscillation) to 1/8 of the total neutrino flux for the case of normal hierarchy, but would gain slightly from 1/3 for the purely oscillating scenario to 2/5 if the neutrino mass eigenstates follow the inverted hierarchy. 

In summary, at the GRB site every UHECR proton that collides with the background GRB or CMB photons would generate 3 neutrinos, among them 2 are muon neutrinos, i.e., the $\nu_{\mu}$ contributes 2/3 of the total population. By the time these neutrinos arrive on Earth, the $\nu_{\mu}$ contribution reduces to 1/3, due to neutrino oscillations. When the neutrino decay is considered, in particular with the assumption of normal hierarchy, then its contribution will become 1/8 when arriving on Earth, a factor 3/8 reduction from that under pure oscillation. On the contrary, if the neutrino mass eigenstates follow the inverted hierarchy, then the $\nu_{\mu}$ population rises from 1/3 to 2/5. We should caution, however, that the notion of neutrino decay is not a direct consequence of the standard model and is not yet experimentally verified.

\section{Detecting GZK Neutrinos with Askaryan Effect}

While the estimate of the GRB induced UHE proton and neutrino fluxes at the GRB site are GRB model and acceleration mechanism dependent, we see that the evolution of the neutrino flavor ratio during their flight to Earth depends on neutrino's fundamental properties. Among them neutrino oscillation is well-established, while neutrino decay still awaits experimental evidence. Based on the discussion in the previous section, however, it should be fair to conclude that, as far as the issue of UHECR cosmic accelerator is concerned, it would be more advantageous if a neutrino observatory can be sensitive to {\it all} flavors. In this regard, an idea proposed by Askaryan in the 1960s (Askaryan \cite{Askaryan1962}; \cite{Askaryan1965})), can in principle detect all neutrino flavors with comparable sensitivities due to the vastness of the typical detection volume involved in such approach. 

As was proposed by Askaryan, high energy cosmic neutrino can be detected by observing the radiowave band of the Cherenkov radiation emitted by the neutrino-induced shower in a large solid target. The neutrino induced shower, though inherently charge neutral, will develop a charge disparity by the time when the shower reaches its maximum, with $\sim 20\%$ more electrons than positrons due to the shorter stopping distance of the positrons propagating in ordinary matter. Since all particles travels near the speed of light, the shower remains compact and thus the emitted Cherenkov radiation is a sharp impulse and therefore is wide-band in the frequency domain, where the radiaowave portion is enhanced due to the coherence of emission. This effect has been validated in a series of experiments in sand, salt and ice performed at SLAC (Gorham \cite{Saltzberg 2001}).

Several ongoing and proposed experiments, e.g., GLUE (Gorham et al. \cite{GLUE2004}), RICE (Kravchenko et al. \cite{RICE2006}), ANITA (\cite{ANITA1-Neutrino}), LUNASKA (James et al. \cite{LUNASKA2010}), ARIANNA (Barwick \cite{ARIANNA2007}; Gerhardt et al. \cite{ARIANNA2010}), and ARA (Chen \& Hoffman \cite{Chen2009}; Allison et al. [ARA Coll.] \cite{ARA2011}), are based on the Askaryan effect. In particular, the balloon-borne ANITA has so far completed two successful missions in Antarctica with exciting results. With an average altitude of 30 km, it tends to detect neutrinos at the higher energy end of the GZK spectrum. So far one candidate neutrino event has been identified from ANITA-2 (Gorham et al. \cite{ANITA-2-2010}). An upgraded ANITA-3 has been scheduled to launch in December 2013 with the expectation of detecting multiple neutrinos. Two new ground-based projects, ARIANNA and ARA, are both at the proof-of-principle stage. For the case of ARA, it is envisioned to cover $200 {\rm km}^2$ area at South Pole with 37 antenna stations. To date (Jan. 2013) 3 stations have been successfully deployed about 200m under ice at Pole. Figure 1 shows the projected ARA37 detector sensitivity in comparison with other projects.

\begin{figure}[htb]
\begin{center}
   \includegraphics[height=8cm]{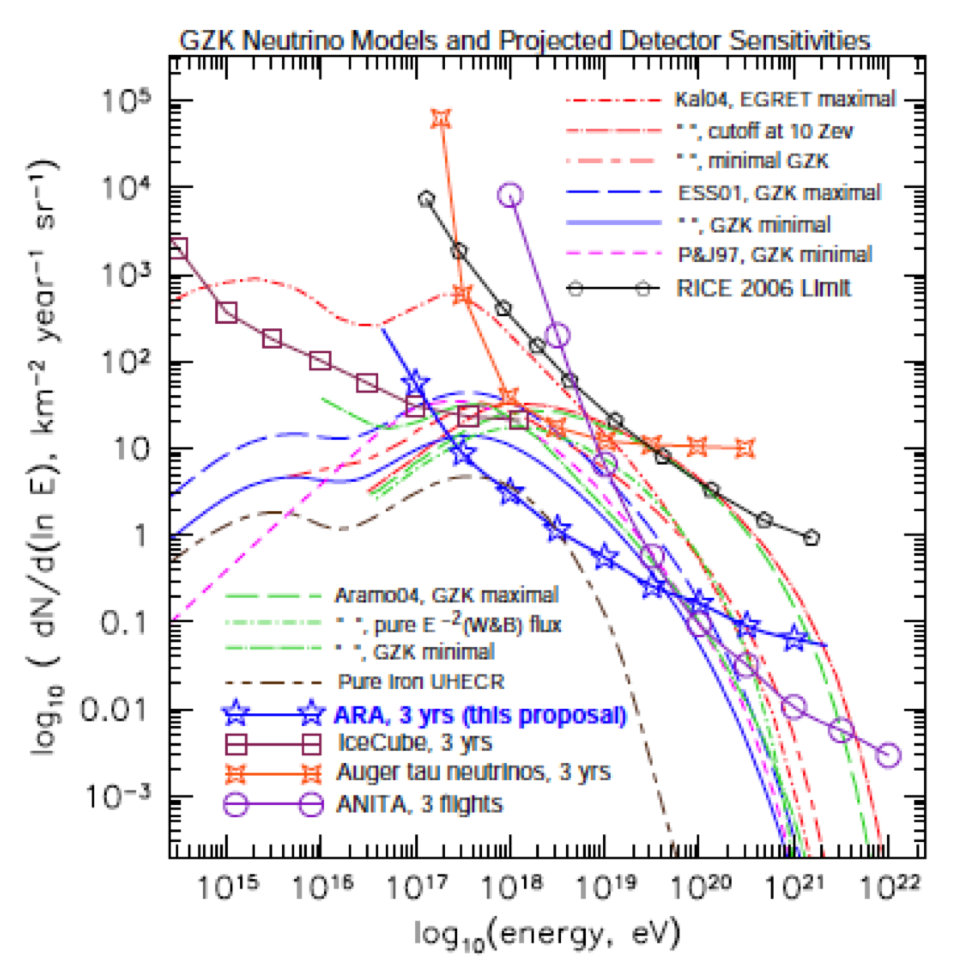}
\end{center}
 \caption{ARA37 sensitivity in comparison with other contending projects plotted against different GZK neutrino models.}
\label{fig:1}
\end{figure}

Focusing on the issue of GRB as a candidate site for UHECR production and its connection to UHE cosmic neutrino detection, what is important for a neutrino telescope is the angular resolution and the {\it total} flux of incoming neutrinos; the composition of neutrino flavors within the total flux does not a priori matter. That said, it would be advantageous if a neutrino telescope for this purpose has indeed the capability of distinguishing neutrino flavors. IceCube can distinguish all three flavors based on different characteristics of their tracks. What about Askaryan effect-based observatories such as ARA? Recent investigations indicate that flavor identification in ARA37 is in principle feasible (Chen, C.-C., Chen, P., Hu et al. \cite{CCChen2013}; Wang, Chen, Huang et al. \cite{SHWang2013}). On the other hand, due to the vastness of the detector's geometric volume and sparseness of antenna stations (2km separation), as a trade-off for its higher sensitivity the angular resolution of Askaryan effect-based observatories tend not to be as good as that for IceCube. For example the angular resolution for ARA37 is $\sim 6^{\circ}$, which is about an order of magnitude worse than that of IceCube muon-neutrino detection. 

 
\section{Conclusion}
We have shown that in addition to the over-estimation by IceCube of the GRB included neutrino flux at-source by a factor $\sim$5 as pointed out by Zhou Li, there are additional impacts to the neutrino flux in-flight as they traverse the cosmos. While the effect due to neutrino oscillations is well-known and has been incorporated into analysis, that due to a possible neutrino decay has received less attention. If neutrinos do decay and their mass eigenstates follow the normal hierarchy, then the net effect is that the flux for muon-neutrino would further reduce from 1/3 under the pure oscillation scenario to 1/8, which would be quite a sizable impact on the IceCube detector sensitivity, assuming that IceCube solely relies on muon-neutrinos in its investigation of the cosmic accelerator question. There are, however, new types of neutrino observatories based on Askaryan effect that are less sensitive to neutrino flavors and would therefore provide almost one order of magnitude improvement in neutrino detection sensitivity. As a trade-off, their angular resolution, on the other hand, will not be as good as that with muon-neutrino at IceCube. If such detector can in addition distinguish different flavors, then the measured flavor ratio would provide a crucial information on the nature of UHECR production, or the inner-workings of cosmic accelerator, at source such as GRB. In conclusion the new type of neutrino observatories based on Askaryan effect maybe complementary to IceCube in shedding lights on this very acute question about the origin of cosmic accelerators.

\section{Acknowledgement}
The author thanks Francis Halzen, Zhou Li, Tsung-Che Liu, and Jiwoo Nam for helpful discussions. This work is supported by Taiwan National Science Council under Project No. NSC100-2119-M-002-025 and by US Department of Energy under Contract No. DE-AC03-76SF00515.


\end{document}